\documentstyle[twocolumn,epsfig,prl,aps]{revtex}
\begin{document}
\draft
\newcommand{\kppp}{\emph{$K_{\tau}$}}
\newcommand{\kdal}{\emph{$K_{Dal}$}}
\newcommand{\ktau}{$K^{+}\rightarrow\pi^{+}\pi^+\pi^-$}
\newcommand{\dal}{$K_{dal}$}
\newcommand{\taus}{$K_{\tau}$}
\newcommand{\sel}{s_e}
\newcommand{\spi}{s_{\pi}}
\newcommand{\thpi}{\theta_\pi}
\newcommand{\thel}{\theta_{e}}
\newcommand{\co}{\; ,}
\newcommand{\po}{\; .}
\newcommand{\nn}{\nonumber \\}
\wideabs{
\title{
A new measurement of $K^+_{e4}$ decay and the $s$-wave
 $\pi\pi$-scattering length $a_0^0$}
\author{S. Pislak$^{7,6}$\cite{SP}, R. Appel$^{6,3}$, 
G.S. Atoyan$^4$, B. Bassalleck$^2$,
D.R. Bergman$^6$\cite{DB}, N. Cheung$^3$, S. Dhawan$^6$, \\
H. Do$^6$, J. Egger$^5$, S. Eilerts$^2$\cite{SE}, W. Herold$^5$, 
V.V. Issakov$^4$, H. Kaspar$^5$, D.E. Kraus$^3$, \\ 
D. M. Lazarus$^1$, P. Lichard$^3$, J. Lowe$^2$, J. Lozano$^6$\cite{JL},
H. Ma$^1$, W. Majid$^6$\cite{WMa}, \\
 A.A. Poblaguev$^4$, P. Rehak$^1$, A. Sher$^3$, 
J.A. Thompson$^3$, P. Tru\"ol$^{7,6}$, and M.E. Zeller$^6$   \\
}

\address{
$^1$ Brookhaven National Laboratory, Upton, NY 11973, USA\\ 
$^2$ Department of Physics and Astronomy, 
University of New Mexico, Albuquerque, NM 87131, USA\\
$^3$ Department of Physics and Astronomy, University of Pittsburgh,
Pittsburgh, PA 15260, USA \\ 
$^4$ Institute for Nuclear Research of Russian Academy of Sciences, 
Moscow 117 312, Russia \\
$^5$ Paul Scherrer Institut, CH-5232 Villigen, Switzerland\\ 
$^6$ Physics Department, Yale University, New Haven, CT 06511, USA\\
$^7$ Physik-Institut, Universit\"at Z\"urich, CH-8057 Z\"urich, Switzerland}
\date{\today}
\maketitle

\begin{abstract}
A sample of $4\cdot 10^5$ events from the decay 
$K^+\rightarrow \pi^+\pi^-e^+\nu_e\;(K_{e4})$ has been collected in experiment
E865 at the Brookhaven AGS. The analysis of these data yields new 
measurements of the $K_{e4}$ branching ratio 
$((4.11\pm0.01\pm0.11)\cdot 10^{-5})$, the $s$-wave $\pi\pi$ scattering 
length $(a_0^0=0.228\pm0.012\pm0.003)$, and the form factors $F$, $G$,
and $H$ of the hadronic current and their dependence on the invariant
$\pi\pi$ mass.
\end{abstract}
}


More than thirty years ago it was recognized that measurements of the
properties of $K_{e4}$ decay $(K^\pm\rightarrow \pi^+\pi^-e^\pm\nu_e
(\bar{\nu}_e))$ would provide important information
about both the weak and strong interactions. This 
four-body semileptonic decay is 
particularly interesting because the 
two pions are the only hadrons in the final state. It allows
studies over a broad kinematic range of several form factors describing
both the vector and axial vector hadronic currents, and
uniquely of the low energy
$\pi\pi$ interaction in an environment without presence of other hadrons.

While experimental studies of $K_{e4}$ held promise of
significant physics insight, the small branching ratio of about 
0.004\% has made precise measurements of the decay parameters
difficult ~\cite{birge65ff,rosselet77}. For instance, while the possibility 
of extracting the 
isospin zero, angular momentum zero scattering length 
$a_0^0$ has long been recognized~\cite{shablin63},
it was not until 1977, when the  
Geneva-Saclay experiment~\cite{rosselet77} gathered about 30,000 events,
that a measurement was made of this quantity to 20\% accuracy.

On the theoretical side, chiral QCD perturbation theory (ChPT)~\cite{ChPT}
makes firm predictions for the scattering length. The tree level calculation 
in ChPT using current algebra
techniques in the soft-pion limit yields 
$a_0^0=0.156$ (in units of $m_\pi$)~\cite{weinberg66}. The one-loop 
($a_0^0=0.201\pm0.01$~\cite{gasser83}) and 
two loop calculations ($a_0^0=0.217$~\cite{bijnens96}) show a satisfactory 
convergence. The most recent calculation~\cite{Colangelo00} 
matches the known chiral perturbation theory representation 
of the $\pi\pi$ scattering amplitude to two loops~\cite{bijnens96} 
with a phenomenological description that relies on the Roy 
equations~\cite{roy71,ananthanarayan00},
resulting in the prediction $a_0^0=0.220 \pm 0.005$.

The analysis of the Geneva-Saclay experiment~\cite{rosselet77} combined with
the Roy equations and the
inclusion of peripheral $\pi N\rightarrow \pi\pi N$ data led to the presently 
accepted value of $a_0^0=0.26\pm0.05$~\cite{nagels79}.
It has been argued, that, if the central experimental value $a_0^0=0.26$
would be confirmed with a smaller 
error, such a large value can only be explained by a significant reduction
of the quark condensate $<0|\bar{u}u|0>$, as is possible in
generalized chiral perturbation theory (GChPT)~\cite{GChPT}.
The quark condensate is an order parameter measuring
the spontaneous breaking of chiral symmetry. Vice versa
measuring $a_0^0$ with higher precision allows to reduce the bounds on this
parameter~\cite{colangelo01}. 

\begin{figure}[htb]
\epsfig{figure=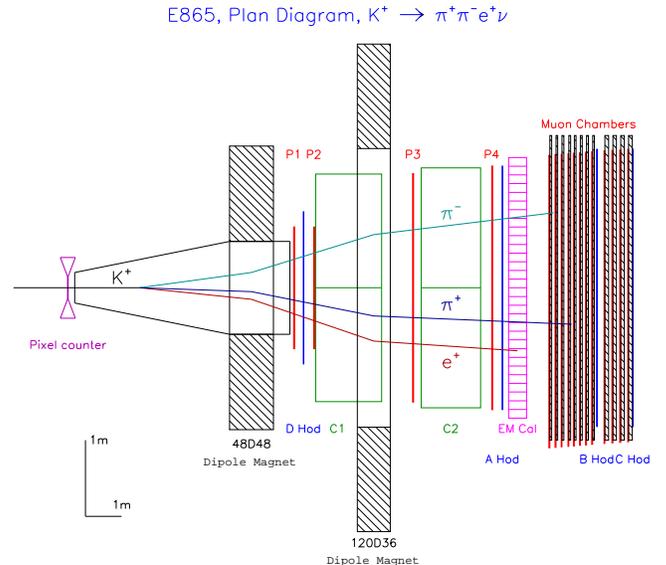,angle=0,width=0.98\linewidth}\centering
\caption[Detector]
{Plan view of the E865 detector. A $K_{e4}$ event is superimposed.
\label{fig:detector}}
\end{figure}

The analysis outlined here is based on data recorded at the 
Brookhaven AGS, employing the E865
detector. The apparatus, described in detail in~\cite{e865},
is shown in Figure~\ref{fig:detector}.
The detector resided in a 6~GeV/c unseparated $K^+$ beam directly downstream 
of a 5~m long evacuated decay volume. A first dipole
magnet separated the $K^+$ decay products by charge.
A second dipole magnet sandwiched between four
proportional wire chambers (P1-P4) served
as spectrometer. 
Two gas \v{C}erenkov counters C1 and C2, filled with CH$_4$ at
atmospheric pressure, and an electromagnetic calorimeter distinguished
 $\pi^\pm$, and $\mu^\pm$ from $e^\pm$. $\pi^\pm$ are separated from
$\mu^\pm$ by a set of 12 muon chambers.
Four hodoscopes were added to the detector for trigger purposes. 
In our analysis we determined the $K^+$ momentum using the beam line as a
spectrometer, the position of the decay vertex and the information from the
pixel counter installed just upstream of the
decay volume.

The first level trigger selected three charged particle tracks 
based on coincidences between the A-, and D-hodoscope and 
the calorimeter. 
The second level trigger indicated the presence of an $e^+$ not
accompanied by an $e^-$. It required signals in both right side and only 
minimal light in both left side \v{C}erenkov counters. In this we 
discriminated against two of the most common background channels: (1) 
$K^+\rightarrow \pi^+\pi^+\pi^-$ ($K_{\tau}$) and (2)
$K^+\rightarrow\pi^+\pi^0$ followed by $\pi^0\rightarrow e^+e^-\gamma$
($K_{dal}$).

The offline analysis selected events containing three charged tracks with
a vertex within the decay volume of acceptable quality, a summed
momentum of less than 5.87~GeV/c, and 
a timing spread between the tracks consistent with the resolution of 0.5~ns.
Even after particle identification criteria were
applied, the remaining sample
still contained background events mainly from
$K_{\tau}$-decay with a misidentification
of a $\pi^+$ as an $e^+$ and accidentals. 
Requiring that the $K^+$ reconstructed from the 
three charged daughter particles does
not track back to the target reduced the background from $K_{\tau}$ to
the level of 1.3$\pm$0.3\%, 
since for $K_{e4}$ the undetected neutrino made
the reconstruction incomplete.
The dominating accidental background was
combination of a $\pi^+\pi^-$ pair from a 
$K_{\tau}$ decay with an $e^+$ from either the beam or 
a coincident decay with an $e^+$ in its final state. A likelihood method
was employed to reduce this background to a level of 2.4$\pm$1.2\%. 
Due to the excellent particle identification capabilities of our detector 
all other backgrounds were negligible.

After the event selection 406,103 events remained, 
of which we estimate $388270\pm5025$
to be  $K_{e4}$ events. 

\begin{figure}[htb]
\epsfig{figure=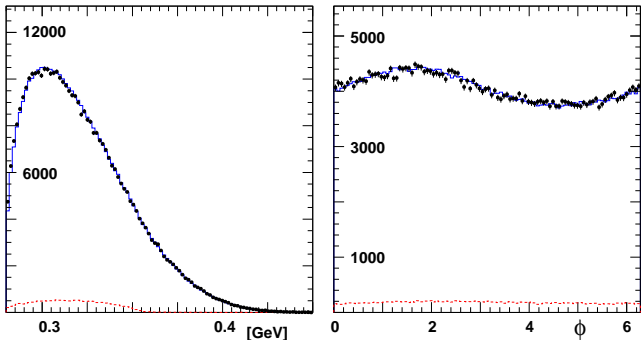,angle=0,width=0.98\linewidth}\centering
\caption[Comparison with Monte Carlo]
{Invariant mass and angle distributions describing the $K_{e4}$ decay. 
The histograms are the Monte Carlo distributions while the
markers with the error bars represent the data. The dashed
histogram indicates the non-$K_{e4}$ background.}
\label{fig:mccompa}
\end{figure}

To determine the branching ratio, the form factors and
other related quantities a Monte Carlo simulation
is needed. Our code, based on GEANT,
takes into account the detector geometry as 
well as the independently measured efficiencies of all detector
elements. $K_{e4}$ decays are modeled by ChPT on the one loop 
level~\cite{bijnens90,riggenbach91}. Radiative corrections 
are included following Diamant-Berger~\cite{diamant76}.
With this apparatus, we generated $81.6\cdot 10^6$ $K_{e4}$ events,
resulting in $2.9\cdot 10^6$ accepted 
events. The agreement between data and
Monte Carlo in all control variable distributions is very good, as e.g.
evidenced by the plots shown in Fig.~\ref{fig:mccompa}.

The $K_{e4}$ branching ratio is measured with respect to $K_{\tau}$ 
decay. $K_{\tau}$ events were collected in a minimum bias prescaled trigger 
together with $K_{e4}$ events. With $Br(\tau)=(5.59\pm0.05)\%$~\cite{groom00}, 
the $K_{e4}$ branching ratio is calculated to be
\[ BR(K_{e4})=\left(4109 \pm 8\,({\rm stat.})\,\pm\,110\,({\rm syst.})
\right)\cdot 10^{-8} \po \]
This result agrees well with the average of 
previous experiments~\cite{groom00}: 
$(3.91\pm0.17)\cdot 10^{-5}$. 
The systematic uncertainties are dominated by the uncertainties
in the \v{C}erenkov counter efficiencies
and background contributions.

The kinematics of $K_{e4}$ decay can be fully described by five 
variables~\cite{cabibbo65}:
(1) $\spi=M_{\pi\pi}^2$, and (2) $\sel=M_{e\nu}^2$, the invariant 
mass squared of the dipion and the dilepton, respectively;
(3) $\thpi$ and (4) $\thel$, the polar angles of $\pi^+$ and $e^+$
in the dipion and dilepton rest frames measured with respect to the
flight direction of dipion and dilepton in the $K^+$ rest frame, respectively;
(5) $\phi$, the azimuthal angle between the dipion and 
dilepton planes.  
The FWHM resolution of the apparatus
for these five variables is estimated to be: 
$0.00133$~GeV$^2$ ($s_{\pi}$), $0.00361$~GeV$^2$ ($s_{e}$),
$147$ mrad ($\theta_{\pi}$), $111$ mrad ($\theta_{e}$), and
$404$ mrad ($\phi$).

The matrix element in terms of the hadronic vector
and axial vector current contributions $V^\mu$ and $A^\mu$ is given by
\begin{table}[htb]
\begin{center} 
\begin{tabular}{lcc}
$M_{\pi\pi}\;(\overline{M_{\pi\pi}})$
& 280-294 (285.2) MeV & 294-305 (299.5) MeV \\
\hline
$F$                   & $5.832\pm0.013\pm0.080$ & $5.875\pm0.014\pm0.083$ \\
$G$                   & $4.703\pm0.089\pm0.069$ & $4.694\pm0.062\pm0.067$ \\
$H$                   & $-3.74\pm0.80 \pm0.18 $ & $-3.50\pm0.52 \pm0.19 $ \\
$\delta$              &$-0.016\pm0.040\pm0.002$ & $0.068\pm0.025\pm0.001$ \\
\hline
$\chi^2/\mbox{NdF}$   & 1.071                   & 1.080                   \\
\hline\hline
$M_{\pi\pi}\;(\overline{M_{\pi\pi}})$ 
& 305-317 (311.2) MeV & 317-331 (324.0) MeV \\
\hline
$F$                   & $5.963\pm0.014\pm0.090$ & $6.022\pm0.016\pm0.094$ \\ 
$G$                   & $4.772\pm0.054\pm0.070$ & $5.000\pm0.051\pm0.082$ \\ 
$H$                   & $-3.55\pm0.44 \pm0.20 $ & $-3.63\pm0.41 \pm0.023$ \\ 
$\delta$              & $0.134\pm0.019\pm0.002$ & $0.160\pm0.017\pm0.002$ \\ 
\hline
$\chi^2/\mbox{NdF}$   & 1.066                   & 1.103                   \\
\hline\hline
$M_{\pi\pi}\;(\overline{M_{\pi\pi}})$ 
& 331-350 (340.4) MeV & $>350$ (381.4) MeV  \\
\hline
$F$                   & $6.145\pm0.017\pm0.096$ & $6.196\pm0.020\pm0.083$ \\
$G$                   & $5.003\pm0.049\pm0.083$ & $5.105\pm0.050\pm0.074$ \\
$H$                   & $-1.70\pm0.41 \pm0.024$ & $-2.23\pm0.48 \pm0.033$ \\
$\delta$              & $0.212\pm0.015\pm0.003$ & $0.284\pm0.014\pm0.003$ \\
\hline
$\chi^2/\mbox{NdF}$   & 1.093                   & 1.034                   \\
\end{tabular}  
\end{center} 
\caption[Fits in individual bins]{Form factors and phase shifts 
$\delta\equiv\delta_0^0-\delta_1^1$ for the six bins in $M_{\pi\pi}$.
The number of degrees of freedom for each fit is 4796. The first uncertainty
is statistical, the second systematical with dominant
contributions from background and \v{C}erenkov efficiency. 
\label{tab:formfactorbin}}
\end{table} 
\begin{eqnarray}
\label{equ:matrixelement}
M=\frac{G_F}{\sqrt{2}} V^*_{us} \overline{u}(p_{\nu})\gamma_{\mu}
(1-\gamma_5)v(p_e)(V^{\mu}-A^{\mu}) \co\\
A^{\mu}=F P^{\mu}+G Q^{\mu} + R L^{\mu}\co\;\;
V^{\mu}=H\epsilon^{\mu\nu\rho\sigma} L_{\nu}P_{\rho}Q_{\sigma}\co 
\end{eqnarray}
where $P=p_1+p_2$, $Q=p_1-p_2$, and $L=p_e+p_\nu$, and $p_1$, $p_2$,
$p_e$, and $p_\nu$ are the 
four-momenta of the $\pi^+$, $\pi^-$, $e^+$, and $\nu_e$ in
units of $M_K$, respectively.  
The form factors $F$, $G$, $R$ 
and $H$ are dimensionless complex functions of $\spi$, $\sel$ and $\thpi$.
The expressions for the decay rate derived from this matrix element 
have been given in ref.~\cite{pais67}.

\begin{figure}[htb]
\vspace*{-8mm}
\epsfig{figure=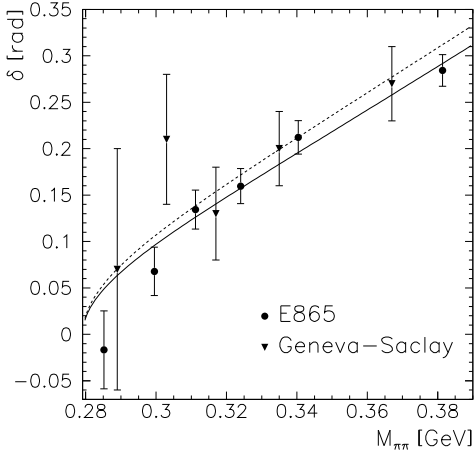,angle=0,width=0.8\linewidth}\centering
\caption[Phase shift difference]
{Phase shift difference $\delta$ as a function of dipion mass. The dashed 
line represents the fit to Eq.~\ref{equ:schenk} for the Geneva-Saclay 
data~\cite{rosselet77} and the solid line for our data as a function 
of the scattering length $a^0_0$.\label{fig:scl}}
\end{figure}

Amor\'os and Bijnens recently developed a parameterisation of these form
factors, based on a partial
wave expansion in the variable $\theta_{\pi}$~\cite{amoros99}:
\begin{eqnarray}
F&=&\left(f_s+f_s^\prime \,q^2+f_s^{\prime\prime}\, q^4+f_e \, s_e\right)
    e^{i\delta^0_0} \nn
 &&+ \tilde{f}_p \,(Q^2/\spi)^{1/2} \,(P\cdot L)\,\cos\theta_\pi \,
    e^{i\delta^1_1} \co\nn
G&=&\left(g_p + g_p^\prime \,q^2 + g_e \,s_e \right) e^{i\delta^1_1},\,
H=\left(h_p+h_p^\prime \,q^2\right) e^{i\delta^1_1} \co
\label{equ:amoros}
\end{eqnarray}     
where $q=(\spi/(4M_\pi^2)-1)^{1/2}$ is the pion momentum in 
$\pi\pi$ rest frame. The form factor $R$ enters
the decay distribution multiplied by $m_e^2$ and can therefore be neglected.
This parameterisation yields ten new form factors $f_s$, $f_s^\prime$, 
$f_s^{\prime\prime}$, $f_e$, $\tilde{f}_p$, $g_p$, $g_p^\prime$, $g_e$,
$h_p$, and $h_p^\prime$, which do not depend on any kinematic
variables, plus the phases $\delta_0^0$ and $\delta_1^1$,
which are functions of $s_\pi$.  

The phase shifts can be related to the scattering lengths. 
A recent analysis~\cite{ananthanarayan00} used
the parameterisation proposed by Schenk~\cite{schenk91}:
\begin{equation}
\tan \delta_\ell^I = \sqrt{1-{4 M_\pi^2 \over s}}\;\sum_{k=0}^3 A^I_{\ell k}
q^{2(\ell+k)}\left({4  M_\pi^2 - s^I_\ell \over s-s^I_\ell} \right) \,.
\label{equ:schenk}
\end{equation}
The Roy equations~\cite{roy71} are then solved numerically, expressing 
the parameters
$A^I_{\ell k}$ and $s^I_\ell$ as functions of the scattering lengths
$a_0^0$ and $a_0^2$. The possible values of
the scattering lengths are restricted to a band in the 
$a_0^0$ versus $a_0^2$ plane. The centroid of this band, the {\em
universal curve}~\cite{morgan69} relates $a_0^0$ and $a_0^2$:
\begin{equation} 
a_0^2 =-0.0849+0.232\, a_0^0-0.0865\, (a_0^0)^2  \po 
\label{equ:universalcurve}
\end{equation} 

For the fits we divided our data
into six bins in $\spi$, five in $\sel$, ten in $\cos\thpi$,
six in $\cos\thel$ and 16 in $\phi$. In the
$\chi^2$ minimisation procedure, the number of measured events
in each bin $j$ is compared to the number of expected events given by:
\begin{equation}
r_j = Br(K_{e4})\frac{N^K}{N^{MC}}\sum \frac{J_5(F,G,H)^{new}}
           {J_5(F,G,H)^{MC}} \co
\label{equ:fitmethod}           
\end{equation}
where the sum runs over all Monte Carlo events in bin $j$. 
$N^K$ is the number of $K^+$ decays derived from the number
of $K_\tau$ events. $N^{MC}$ is the number of generated events. 
$J_5(F,G,H)^{MC}$ ($\equiv I$~\cite{pais67}) is the five-dimensional phase 
space density
generated at the momentum $q=q^{MC}$ with the form factors $F$, $G$, 
and $H$ used to simulate the event.
$J_5(F,G,H)^{new}$ is calculated at $q^{MC}$ with $F$, $G$, $H$ evaluated
from the parameters of the fit. 
Thus, we apply the parameters on an event by event basis,
and, at the same time, we divide out a possible bias caused by the matrix
element, making the fit independent of the ChPT ansatz 
used to generate the MC. 

For the fit, we have assumed that $F$, $G$, and $H$ do not depend
on $\sel$ and that $F$ contributes to $s$-waves only, 
i.e. $f_e=g_e=\tilde{f}_p=0$.
Our first set of fits is done 
independently for each bin in $\spi$.
The above assumptions then
leave one parameter each to describe $F$, $G$, and
$H$ aside from the
phase difference $\delta\equiv\delta_0^0-\delta_1^1$.
The results are listed in Table~\ref{tab:formfactorbin}.
The centroids of the bin ($\overline{M_{\pi\pi}}$) 
are determined following Lafferty
and Wyatt~\cite{lafferty95}. If the six phase shifts in 
Table~\ref{tab:formfactorbin} are fit using
Eq.~\ref{equ:schenk} and Eq.~\ref{equ:universalcurve},
one obtains $a_0^0=0.229\pm0.015\;\;(\chi^2/\mbox{NdF}=4.8/5$). The 
resulting curve is shown in Fig.~\ref{fig:scl}.
\begin{table}[htb]
\begin{center} 
\begin{tabular}{lrlr}
$f_s$                 & $ 5.75\pm0.02 \pm0.08$   
& $f_s^\prime$          & $ 1.06\pm0.10 \pm0.40$  \\ 
$f_s^{\prime\prime}$  & $-0.59\pm0.12 \pm0.40$   
& $g_p$                 & $ 4.66\pm0.05 \pm0.07$  \\ 
$g_p^\prime$          & $ 0.67\pm0.10 \pm0.04$   
& $h_p$                 & $-2.95\pm0.19 \pm0.20$  \\ 
\multicolumn{4}{c}{$a^0_0$\hspace*{10mm} $0.228\pm0.012\pm0.004$}\\
\end{tabular}  
\end{center} 
\caption[Mass dependent fit]{Form factors and scattering length $a^0_0$
in the parameterisation of Eq.~\ref{equ:amoros} 
($\chi^2/\mbox{NdF}=30963/28793$.)}
\label{tab:ffwithuniversal}
\begin{center} 
\begin{tabular}{lrc}
$\tilde{f}_p$         & $ -4.3\pm1.3 \pm\;\;3.4$  & 
$\chi^2=30952$\\ 
$f_e$                 & $ -4.1\pm1.3\pm\;\;3.1$  & 
$\chi^2=30954$ \\ 
$g_e$                 & $  0.5\pm4.4\pm11.3$ & 
$\chi^2=30963$\\ 
\end{tabular}  
\end{center} 
\caption[]{Fit of form factors $\tilde{f}_p$, $f_e$, and $g_e$. The 
systematic uncertainties are dominated by the resolution of the neutrino
mass squared. (NdF=28792)}
\label{tab:ffhigerorder}
\end{table}

We have also made a single fit to the entire data sample. In this
second fit we 
substituted $\delta$ in 
Eq.~\ref{equ:amoros} by the expression of Eq.~\ref{equ:schenk}.
With the relation between $a^0_0$ and $a^2_0$ given
by Eq.~\ref{equ:universalcurve} only
$f_s$, $f_s^\prime$, $f_s^{\prime\prime}$, $g_p$, $g_p^\prime$,
$h_p$, and $a_0^0$ then remain as free parameters. 
The results, 
listed in Table~\ref{tab:ffwithuniversal}
are in an excellent agreement with the ones derived in the
previous paragraph.
 
To check the assumption $f_e=g_e=\tilde{f}_p=0$
we also allowed these form factors to vary, one at the
time, in our second fit. Table~\ref{tab:ffhigerorder} shows that all 
three form factors are consistent with zero.
The quality of the fits is demonstrated in
Fig.~\ref{fig:mccompa}, where the invariant mass 
($\spi$) and azimuth ($\phi$) distribution from data
are compared to the reweighted Monte Carlo distributions
(Eq.~\ref{equ:fitmethod}). 
The asymmetry of the $\phi$
distribution is the only observable, which directly depends
on the phase shifts~\cite{pais67}. The amplitude of this asymmetry 
amounts to only 10\% of the $\phi$ independent part. This explains
why the statistical error of $a^0_0$ is still
limited to 5.3\%.

To summarize,
experiment E865 has collected a 
$K_{e4}$ event sample more than ten times larger 
than all previous experiments combined. From the
model independent analysis of this data the momentum dependence
of the form factors of the hadronic currents as well as $\pi\pi$ scattering
phase shifts have been extracted. The form factors and phase shifts serve
as an important input in the program to determine
the couplings of the effective Hamiltonian of chiral QCD perturbation theory
at low energies~\cite{talavera00}. From a preliminary communication
of these results already tight bounds on the value of the
quark condensate have been extracted~\cite{colangelo01}.
Using the relations between $a_0^0$ and $a_0^2$ given by the Roy 
equations~\cite{ananthanarayan00}, we have extracted the most precise
value of the $\pi\pi$ scattering length $a_0^0$. This 
value agrees well with predictions obtained 
in the framework of ChPT~\cite{Colangelo00}.

We gratefully acknowledge the contributions to the success of
this experiment by Dave Phillips,  
the staff and management of the AGS at the Brookhaven National
Laboratory, and the technical staffs of the participating institutions.
We also thank C.~Colangelo and J.~Gasser for many
fruitful discussions and helpful comments.
This work was supported in part by the U. S. Department of Energy, 
the National Science Foundations of the USA, Russia and Switzerland, and
the Research Corporation.

\end{document}